# Asymmetry of Endofullerenes with Silver Atoms

V.S. Gurin

*Physico-Chemical Research Institute, Belarusian State University*

*Leningradskaja str., 14, Minsk, 220080, BELARUS;*

*E-mail: gurin@bsu.by; gurinvs@lycos.com*

**Abstract**

A series of endofullerenes Ag@$C_{60}$ with different symmetry are calculated at *ab initio* level. The lowest energy structure is completely asymmetrical one ($C_1$), in which the endo-atom has noticeably off-centre position. The symmetrical structures are less stable. Silver atom in the Ag@$C_{60}$ ($C_1$) endofullerene has the low negative charge and high spin density.

## 1. Introduction

Endohedral fullerenes, $M_x$@$C_{60}$, are of interest as species in which interaction of metal and carbon atoms occurs in the unique manner since an atom is confined by the carbon cage [1,2]. The geometrical confinement effect may not be *a priori* combined with a chemical bonding, though in the case of atoms of typical metals inside $C_{60}$ (and higher $C_n$) the bonding is really takes place [3,4]. The endofullerenes are rather different both in chemical and physical properties as compared with corresponding exo-counterparts as well bare metal atoms and $C_{60}$ molecules without metals. An atom inside of the fullerene cage can provide a profound effect upon electronic properties and geometry, and both M atom and $C_{60}$ form new chemical species. The stable $C_{60}$ molecule with closed electronic shell is not very active reagent for metal atoms, while the higher fullerenes $C_{70}$, $C_{82}$, $C_{84}$, etc. are more reactive. A rich diversity of endostructures exists with the higher fullerenes, however the problem with $C_{60}$ is of special interest in connection with high symmetry of the bare $C_{60}$ molecule and its easier availability in production to design some novel materials and advanced devices.

More frequent cases of endofullerenes are known with active metals (alkali, earth-alkali and rare earth elements) [1-4], however, many other atoms may exist inside fullerene cages also. Silver as possible endoatom has been proposed in few recent works [5,6], though an experimental confirmation was not found to date yet to our best knowledge. However, it is not only academic interest can be for the extension of the circle of possible metals inside $C_{60}$, but also their physical properties are expected to be of interest since metal atoms and clusters those are instable in environment can be kept within fullerene cages. We emphasize here the spin properties of $Ag@C_{60}$ and study the effect of asymmetry in geometry of these species. The asymmetry of $M@C_{60}$ which occurs for the cases with various metals appears to be rather challenged feature of endofullerenes.

The principal concept of bonding of metals within fullerenes is the electron donation mechanism [3]. Chemical elements with easy capability to produce endohedral fullerenes commonly have the low ionization energy, that is typical for atoms of active metals. The typical non-metals with no pronounced electron-donor ability, e.g., nitrogen, phosphorous, rare gases can also produce stable endofullerenes [7], in which an interaction of endoatoms with the carbon cage is weak. Silver has the ionization energy 7.6 eV that is very close to the known value of $C_{60}$ [8] and does not fit the above rule for active metals. Meanwhile, the geometrical factor of confinement to produce $M@C_{60}$ with M=Ag remains to be significant, and size of silver atoms does not provide any restriction for ~7 Å $C_{60}$ cage. The high symmetry of molecule $C_{60}$ ($I_h$), at the first sight, proposes any single atom located in the symmetry centre of the icosahedron that is correspond to this $C_{60}$ geometry. However, this intuitive prediction did not true in general case, and in the present work we study the deviations for $Ag@C_{60}$ models. For the cases of another elements those also provide asymmetrical (off-centre) position of endoatoms, Na, Li, Cu, Be, etc, the situation with asymmetry effect can be qualitatively similar and will be studie in future.

## 2. Construction of Models and Calculation Method

For $C_{60}$ endofullerenes we consider the models of different symmetry putting an endoatom at the center, but the symmetry group under calculation was $I_h$ and the subgroups allowed the off-centre shift of M (Table 1): (i) the model with $I_h$ symmetry describe perfect symmetrical endofullerene, and geometry optimization may vary only C-C distances; (ii) the models with the lower symmetry, $C_2$, $C_3$, $C_5$, allowing a distortion of an original geometry keeping the corresponding symmetry to attain a minimum of energy, and (iii) full asymmetrical model ($C_1$ point group). These cases really mean the deviations of the endoatom from the central position to any direction ($C_1$) or along the symmetry axes ($C_2$, $C_3$, $C_5$). The case (i) does not allow any shift of endoatom. There are no a priori information on preference some of the above symmetry reduction, but preliminary calculations showed that $C_1$ distortion from $I_h$ is quite probable with silver endoatoms. Fig. 1 depicts the undisturbed fullerene molecule and these axes, and Fig. 2 present the geometries of Ag@$C_{60}$ calculated for the above cases of the off-centre shift along the axes or an arbitrary shift (iii).

Together with the possible shifts of silver endoatom, a little deformation of original C-C bonds in $C_{60}$, evidently, can take place as a result of geometry optimization with the lower symmetries. The corresponding coordinates of carbon atoms to describe initial geometry of $C_{60}$ were generated from the calculation data for $C_{60}$ with $I_h$ symmetry, and the same ones were taken in the case of $C_5$ symmetry. The $C_3$ model was obtained by the rotation at 36 degrees from the main axis and $C_2$ one was done by 90 degrees rotation.

We use the SCF Hartree-Fock methods within MOLCAO (molecular orbitals – linear combination of atomic orbitals) approach with full geometry optimization within the given symmetry groups. The ground state of $C_{60}$ is known to be a singlet, and for Ag@$C_{60}$ we consider doublet states calculated within the unrestricted Hartree-Fock method. The doublets corresponded to the lowest energy in these models. The basis functions were constructed with the 19-electronic effective core potential (ECP) for Ag and the all-electronic set of STO-3G and

6-31G quality for carbon atoms. The calculations were done with a NWChem 4.1-4.5 software [9] and the basis function were used according to data within the package.

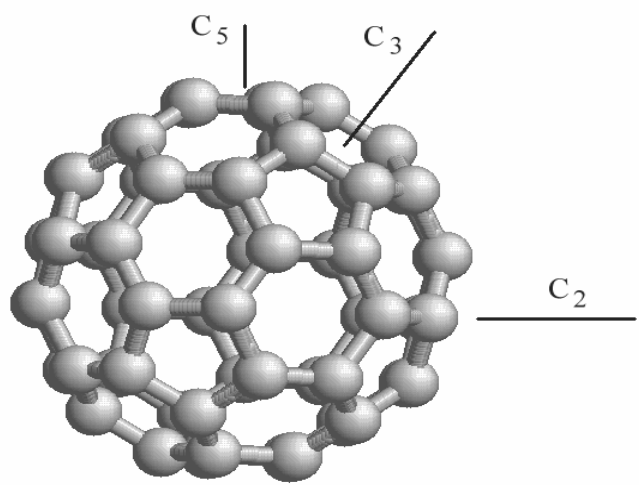

Fig. 1. Geometry of the fullerene $C_{60}$ molecule without endoatoms with the symmetry axes

Table 1. Calculated data for a series of Ag@$C_{60}$ models of different symmetry

| Model and symmetry | Min C-C Å | Max C-C Å | Off-centre shift in Ag@$C_{60}$, Å | Binding energy of Ag@$C_{60}$, eV | Effective charge of Ag, e | Spin density at Ag |
|---|---|---|---|---|---|---|
| $C_{60}$ ($I_h$) | 1.375 | 1.463 | | | | |
| Ag@$C_{60}$ ($I_h$) | 1.40 | 1.54 | 0 | <<0 | 0 | 0 |
| Ag@$C_{60}$ ($C_1$) | 1.40 | 1.54 | 0.28 | 9.94 | -0.10 | 0.99 |
| Ag@$C_{60}$ ($C_2$) | 1.39 | 1.54 | 0.05 | 4.55 | -0.92 | 0.01 |
| Ag@$C_{60}$ ($C_3$) | 1.30 | 1.61 | 0.011 | -13.97 | -0.94 | 0.0 |
| Ag@$C_{60}$ ($C_5$) | 1.43 | 1.55 | 0.008 | -4.08 | -1.89 | 1.03 |

**3. Results and Discussion**

The calculation results are collected in Table 1 and used to display the geometry of models in Fig. 2. The most featured result of these calculations is the strong dependence of stability of Ag@$C_{60}$ structures on symmetry conditions. The binding energy we have defined as the difference between full electronic energy of Ag@$C_{60}$ and the separated constituents, Ag and $C_{60}$, calculated with the same parameters. The positive binding energy means stability, the lower electronic energy of the bound structure, and negative one corresponds to instability. This value appeared to be of maximum for completely asymmetrical model, $C_1$, and this value is rather high (~10 eV), i.e. an existence of Ag@$C_{60}$ ($C_1$) may be expected at room temperature. Recently [10], Cu@$C_{60}$ endofullerene was shown to be produced by collision of evaporated $C_{60}$ molecules with copper plasma. As far as properties of copper and silver are rather close in properties we may suggest possibility of similar production of Ag-containing endofullerene. Cu@$C_{60}$ has been calculated at the same theory level [6], thus, predicting the experimental data in Ref. 10.

The model of $C_2$ symmetry demonstrates the less stability and stronger distortion of the carbon cage. The structures of the higher symmetry, $C_3$, and $C_5$ are instable, however, they possess the lower energy than the model with the perfectly central position of silver since the geometry optimization has been done taking into account possible positions of Ag along the corresponding axes. These models show also very high value of effective charge at Ag, about 1 e that looks rather anomalously. Meanwhile, in the case of $C_1$ model the effective charge is quite reasonable, slightly negative, -0.1 e. The negative charge can be explained if to remember the fact that the ionization potential of Ag is close to that for $C_{60}$ (noted in Introduction). In the case of the other metal atom inside, like Na, Mg, the effective charge appear to be positive. Thus, the charge transfer between Ag and $C_{60}$ cage in the asymmetrical Ag@$C_{60}$ is not strong and reversed as compared with active metals.

One of interesting aspect of endofullerene geometry is off-centre position of endoatoms. In these calculation results we can note that the maximum shift corresponds to the most stable structure. Moreover, this shift takes place not along any symmetry axis (the displacement value for three x,y,z coordinates, and the square root from the sum of $\Delta$x, $\Delta$y and $\Delta$z is given in Table 1.

The structure Ag@$C_{60}$ with $C_1$ symmetry if featured the high spin density at silver atom, while the other models (besides $C_5$) indicate the low value for spin density. The latter is also zero for $I_h$ structure. Thus, the asymmetrical model can be proposed as spintronic element in construction of quantum devices [11]. The spin concentrated at silver atom stabilized by the case in the endofullerene, however, the magnetic features of this model can be managed externally.

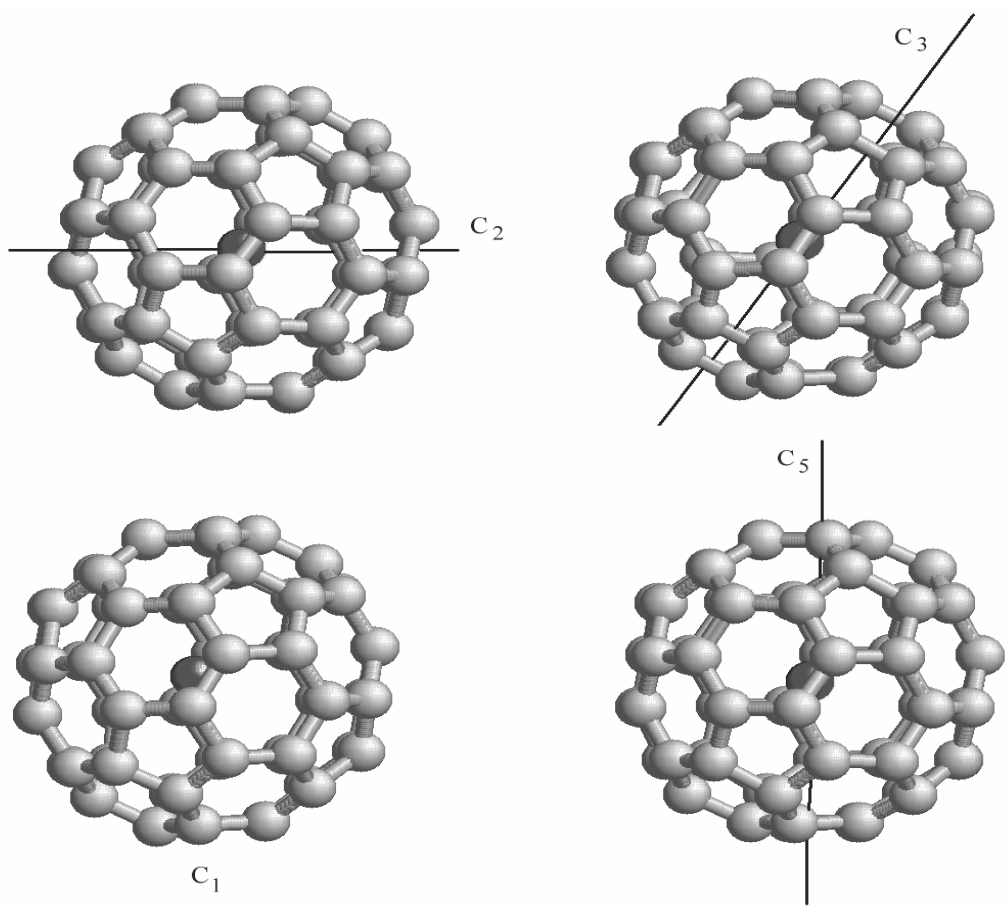

Fig. 2. Geometry of four AgC$_{60}$ models calculated with different symmetry

**Acknowledgments**

The author acknowledges the support of this work due to participation in the project under the Ministry of Education of Belarus